\documentclass[journal=jpclcd,manuscript=letter,biblabel=brackets]{achemso}
\usepackage{fontenc}
\usepackage{graphicx}
\usepackage{amssymb,amsmath}
\usepackage{caption}
\usepackage[version=3]{mhchem}

\date{\today}
\date{}
\author{Marcel D.~Baer}
\affiliation{Chemical and Materials Science Division, Pacific Northwest National Laboratory, Richland}
\author{Abraham C.~Stern}
\affiliation[University of California, Irvine]
{Department of Chemistry, University of California, Irvine, 
        Irvine, California 92697-2025, United States}

\author{Yan Levin}
\affiliation{%
Instituto de F\'isica, UFRGS
Porto Alegre, RS, 91501-970,
Brazil
}
\email{levin@if.ufrgs.br}
\author{Douglas J.~Tobias}
\affiliation[University of California, Irvine]
{Department of Chemistry, University of California, Irvine, 
        Irvine, California 92697-2025, United States}
\email{dtobias@uci.edu}
\author{Christopher J.~Mundy}
\affiliation{Chemical and Materials Science Division, Pacific Northwest National Laboratory, Richland}
\email{chris.mundy@pnnl.gov}

\title{%
The Electrochemical Surface Potential Due to Classical Point Charge Models Drives Anion Adsorption to the Air-Water Interface } 

\begin{document}
\begin{abstract}
We demonstrate that the driving forces for ion adsorption 
to the air-water interface for point charge models 
results from both cavitation and a term that is of the form of 
a negative electrochemical surface potential.  
We carefully characterize the role of the free energy
due to the {\it electrochemical} surface potential 
computed from simple empirical models and its role in ionic adsorption within the
context of dielectric continuum theory. 
Our research suggests that the electrochemical surface potential 
due to point charge models
provides anions with a significant driving force to the air-water interface.
This is contrary to the results of {\it ab initio} simulations that indicate
that the {\it average electrostatic } surface potential should favor the desorption of anions
at the air-water interface.   The results have profound implications for the studies
of ionic distributions in the vicinity of hydrophobic surfaces and proteins.
\end{abstract}
\begin{scheme}
\includegraphics[width=5.1cm]{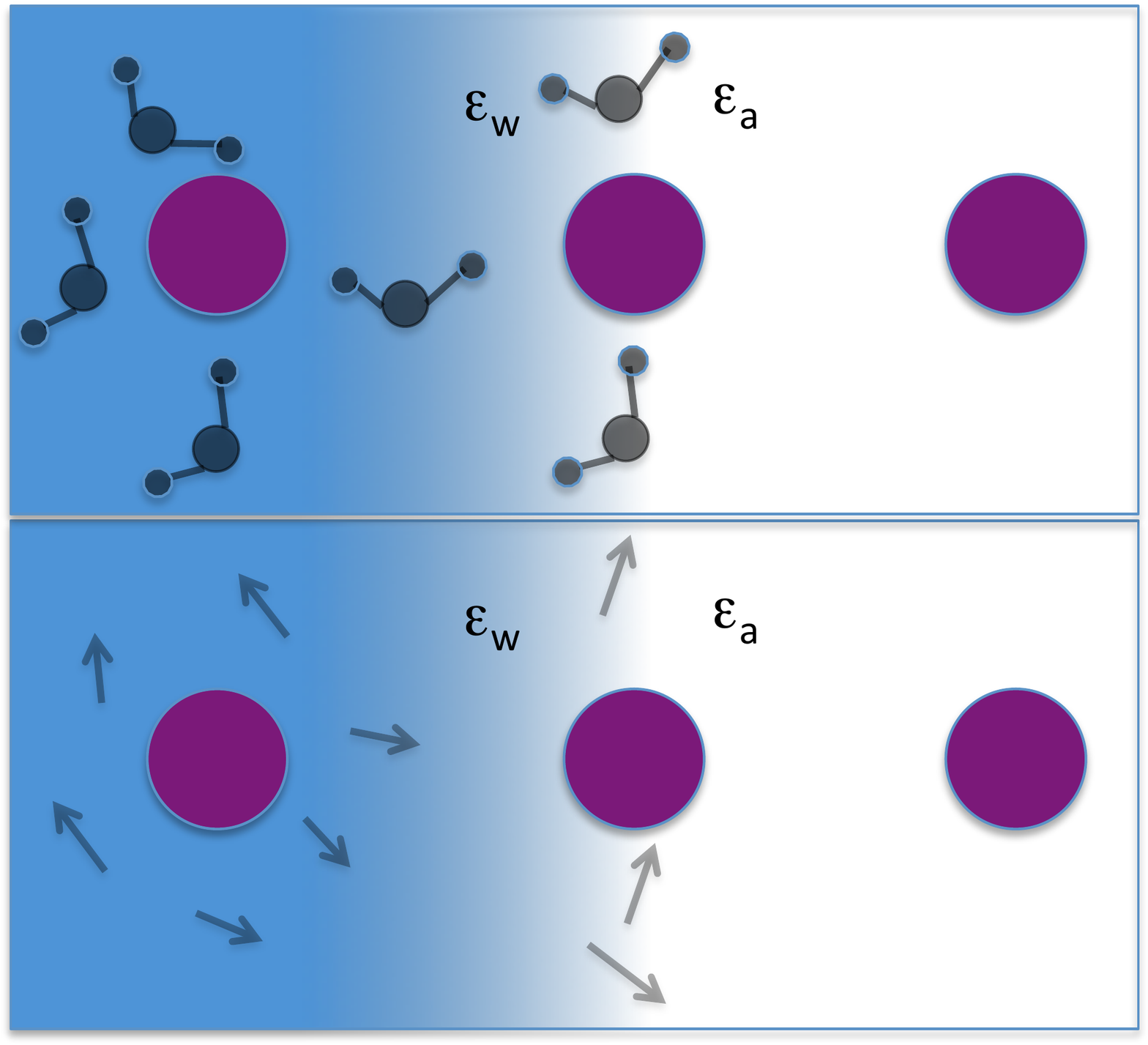}
\end{scheme}

\newpage

Wolfgang Pauli used to say ``God made the bulk; the surface was invented by the
devil.''~\cite{Jamtveit1999}  Pauli's frustration referred to solid surfaces, which are relatively
simple compared to liquid interfaces.  Besides the broken symmetry associated
with the solid surface, liquid-liquid and liquid-air interfaces exhibit capillary fluctuations, 
making their study even more difficult.  If one of the liquid phases 
contains a mixture --- solvent plus solute  --- the level of complexity deepens.  
We can only speculate about what Pauli would say about such a confounding geometry.

The first clues to the behavior of ions near the air-water interface came from the
measurements of excess surface tension of electrolyte solutions.~\cite{Heydweiller1910}
In particular, it was observed that halide salts increase the surface tension of the air-solution interface. 
For this to be the case, thermodynamics requires that ions must be depleted from the 
interfacial region.~\cite{Onsager1934}
On the other hand, electrochemical measurements of Frumkin showed that the electrolyte solutions possess
an electrostatic potential gradient across 
their surface.~\cite{Frumkin1924} Frumkin's work clearly indicated that cations
and anions can behave very differently near the surface by demonstrating that anions are closer to the
surface than cations.  Frumkin's work was also consistent with another more than 100 year old
mystery: the Hofmeister series of electrolyte solution.~\cite{Frumkin1924} In 1888 
Hofmeister observed that various monovalent electrolytes have very different effects on
protein solutions.~\cite{Kunz2004} While some electrolytes help solubilize 
proteins, denaturing them in the process,
others lead to protein precipitation.  
The interaction of ions with surfaces, such as 
water-protein or air-water interfaces, has remained an outstanding puzzle of physical chemistry.  

Indications to the mechanism of the 
ion-surface interaction started to appear when computational methods, hardware and
software became sufficiently powerful to allow realistic microscopic simulations. 
The work of Berkowitz on small water clusters in the early 1990s~\cite{Perera1993} and 
of Jungwirth and Tobias (JT) on extended
air-water interface a decade ago brought to the
attention of the physical chemistry community a 
new phenomenon associated with aqueous electrolytes,~\cite{Jungwirth2001} namely,
that soft, polarizable anions can adsorb to the 
air-water interface. 

Levin and co-workers have recently shown that it is possible to quantitatively explain the surface
tension measurements and the surface adsorption of large polarizable anions in 
terms of a dielectric continuum theory that explicitly takes into account the finite
size and a full polarization response of the ions comprising the electrolyte to the 
presence of the interface.~\cite{Levin2009,Levin2009a,Santos2010}
By using the accepted literature values of anionic radii and polarizabilities,
the radius of the hydrated sodium counter ion was fit to quantitatively reproduce the surface tension
data of the alkali halide and other oxyanion salts.  The results of this polarizable anion
dielectric continuum theory (PA-DCT) were qualitatively consistent with the 
original work of JT.

An important difference between PA-DCT theory and JT is that the 
predicted amount of adsorption for the most polarizable
iodide anion is significantly less in the former.  
State-of-the-art {\it direct} surface measurements 
suggest that there is indeed adsorption of large polarizable anions to the air-water 
interface.~\cite{Ghosal2005,Petersen2006,Baer2009}
However, precise measurements on the degree of anion adsorption at low concentrations usually
rely on using the Gibbs' adsorption isotherm (GAI) that relates surface excess of anions 
to the electrolyte concentration dependence of the surface tension.~\cite{Ho2003}  
For the case of iodide, the large adsorption free energies predicted by JT would produce 
concentration dependent surface tensions that resemble surfactants rather than electrolytes.

Although quantitatively different, both JT and the PA-DCT approaches
point to the fundamental role of ionic polarizability for ion adsorption.
Recently, a potential of mean force (PMF) of iodide at the air-water interface 
was computed using molecular 
interactions based on the density functional 
theory (DFT) (a so called {\it ab initio} simulation that naturally takes
into account the full polarization response of the anion to the air-water interface)
and yielded results in almost quantitative agreement with the PA-DCT.~\cite{Baer2011}

In a recent study, Horinek 
and co-workers have fit a {\it non-polarizable} point charge, 
soft-sphere interaction potentials for the alkali-halide salts in a fixed charge
{\it non-polarizable} water model. The resulting PMF obtained from the simulation was fitted 
to reproduce the experimentally 
measured surface tensions as a function of 
electrolyte concentration.~\cite{Horinek2009}   
Interestingly, the PMF of a 
single iodide anion from the {\it ab initio} calculation and the PA-DCT compare 
well to the work of Horinek and co-workers: all of the PMFs contain a shallow minimum that is less
than $k_{\rm B} T$, where $k_{\rm B}$ is Boltzmann's constant and $T=300$ K. 
This is to be contrasted to the PMFs of iodide obtained with a polarizable force field similar to 
those used in the original study of JT that predicts the surface 
adsorption of roughly $3k_{\rm B}T$.~\cite{Dang2002,Baer2011}
The results of the fitted {\it non-polarizable} force fields of
Horinek and co-workers~\cite{Netz2012,Arslanargin2012}
provide additional validation of the depth of the
PMF minimum at the surface that is necessary to be consistent with surface tension
measurements.  However, other questions regarding the role of polarization in
the driving force of anions to interfaces remain unanswered. 

The purpose of this paper is to show that the driving force
for adsorption of large halide anions when using {\it non-polarizable} point
charge models for both the anion and water is due to very different physics than 
 the {\it ab initio} MD simulation and PA-DCT.
We do not dispute the success of point charge models 
({\it e.g.} SPC/E)  for describing a range of both bulk and 
interfacial properties of neat water.~\cite{Paschek2004,Kuo2006} 
However, clues to a possible model dependence for the 
driving force of anions to interfaces may lie in 
the surface potential of the neat air-water interface that could
strongly influence ionic adsorption.~\cite{Arslanargin2012}
The surface potential of the air-water interface computed utilizing a classical point 
charge model, $-0.6\,\rm{V}$ differs in both sign and magnitude
to the {\it ab initio} result of $+3\rm{V}$.~\cite{Kathmann2008,Kathmann2009,Kathmann2011}  
In this study we will make a distinction between the electrochemical surface 
potential --- the potential of mean force that
an ion experiences due to the presence of an interface --- and the average
electrostatic  potential due to the presence of the interface, which is the surface potential.

The PA-DCT is based on the assumption that there is {\it no significant} 
free-energy contribution from the {\it electrochemical} surface potential
across the neat air-water interface.  This ansatz is corroborated
by the good agreement of the computed electrochemical surface potentials 
with PA-DCT and the measurements of Frumkin on electrolyte solutions 
which are indeed small (on the order of $\rm{mV}$).~\cite{Frumkin1924} 
An important conclusion that arises from the comparison PA-DCT and {\it ab initio} simulations is that 
ions in the DFT approximation, for reasons that are not completely clear at
this time,  feel an electrochemical surface potential of the air-water
interface that is nearly zero.~\cite{Baer2009,Mundy2009,Baer2011}  

In this letter we will isolate the effect of the electrochemical surface potential 
on driving ions to interfaces 
by examining the solvation of a hard-sphere ion in both an SPC/E
model of water and a Stockmayer fluid, which by symmetry has no surface
potential.  Through the choice of a hard-sphere solute we can  
cleanly isolate the  volume dependent cavitational energy 
from the dielectric response of the charging process.
Thus, we will demonstrate that the  models presented herein provide a 
straightforward decomposition of the ion adsorption propensity 
into cavitation and self-energy.
The tug-of-war between the cavitation and self-energy penalty usually favors the bulk ionic solvation
for {\it non-polarizable} models of anions.
However, we will show that for fixed point charge models of ions and water there 
is an additional contribution to the free
energy due to the electrochemical surface potential of the air-water interface.  
From these results two important findings will emerge: 
first, it will be shown that the free energy due to the electrochemical
surface potential that a charged hard sphere feels as it approaches the interface is not 
the average surface potential of the air-water interface as is 
normally computed~\cite{Wick2009a,Leung2010,Kathmann2011,Arslanargin2012}.  
Second, we will demonstrate that the principles of anion adsorption are
model dependent.  The fundamental role of the electrochemical surface potential
as the driving force of ions to interfaces found in point charge models 
is inconsistent with the mechanism 
that arises when a quantum mechanical description of the charge density is
utilized.


An extended interfacial system containing 215 solvent molecules and 
a single iodide anion was modeled in slab geometry within a supercell of
15$\times$15$\times$75\,\AA$^3$. This choice of system
size has been shown to produce a stable bulk liquid in the center of the
slab and is able to quantitatively reproduce structural properties of larger system
sizes that have been used elsewhere.~\cite{Kuo2006}

It is clear from our results (see the supporting information (SI)) that we have a stable liquid for
the Stockmayer fluid that can be used to embed a charged hard sphere cavity.  
Moreover, the surface potentials
for both the Stockmayer and SPC/E fluid can be computed using the simple formula first presented by Wilson {\it et al}.~\cite{Wilson1988}
The resulting surface potential for SPC/E water is in excellent agreement with all studies performed on this
system to date yielding a value of roughly $-0.6$ V and validating our choice of system size.  
As expected, due to symmetry, the Stockmayer fluid yields a
value of approximately $0$ V.  
In order to make contact with a dielectric continuum theory  we estimate the dielectric constant of both 
SPC/E water and the Stockmayer fluid using the same number of solvent molecules in a cubic box at the estimated
bulk density.~\cite{Spoel1998} The dielectric constant for the SPC/E (65) is in good agreement with the reported values
for 216 water molecules by van der Spoel \textit{et al.} (69),~\cite{Spoel1998} as our estimated bulk density is a bit lower.

Thus, we are positioned to compare the statistical mechanics of ion adsorption of a 
dielectric medium with zero and finite values of the surface potential.
The dielectric continuum theory (DCT) presented herein, is built by considering various contributions 
to the solvation free energy of a hard-sphere ion (which differs from the PA-DCT discussed above).
We first note that to transfer an ion into water requires the creation of a cavity.  For small cavities, less than 4~\AA,
the cavitational energy is proportional to the volume of the cavity.  When the ion of charge $q$ moves across the interface,
the cavitation energy decreases proportionally to the volume exposed to the vapor phase.  The
cavitation energy for an ion of radius $a$ located at distance $z$ $\left[-a,a\right]$ from the Gibbs dividing surface (GDS)~\cite{Jungwirth2006} (positive $z$ is towards water)
is given by:
\begin{equation}
\beta F_{\rm cav}(z)=0.075 a^3 \left( \frac{z}{a}+1\right)^2 \left( 2-\frac{z}{a}\right)
\label{eq:cav1}
\end{equation}

\noindent where $F_{\rm cav}$ is in units of $k_{\rm B} T$ and the prefactor is in units of \AA${}^{-3}$.~\cite{Levin2009}
When the ion is in the bulk water, its electrostatic field is screened by the surrounding water molecules so that 
the electrostatic self-energy is $\beta F_{\rm self}=\lambda_B/2 a$, 
where the Bjerrum length is $\lambda_B=e^2/\epsilon_{\rm w} k_{\rm B} T$.  
When the ion is at distance $z=a$
from the interface, besides its interaction with the water molecules, it also interacts with the induced charge
on the dielectric interface.  The electric field produced by the induced charge is exactly the same as the
field produced by a point image charge located across the interface.  Since the dielectric contrast between water and
air is so large, the image charge has the same sign and magnitude as the ion, and is  located at 
$z=-a$.  The electrostatic self energy of an ion located at $z=a$ can then be easily calculated to be
$\beta F_{\rm self}(a)=3 \lambda_B/4 a$.  When the ion crosses the interface and is at $z=-a$,
it sees an image in water, which has the charge $-q$, and the self energy is therefore 
$\beta F_{\rm self}(-a)= \epsilon_{\rm w} \lambda_B/4 \epsilon_{\rm air} a$.  Unfortunately, there is no simple way
to calculate the self energy for an ion that has only partially penetrated the interface; the energy will
depend on the precise model that is use to treat the interior of an ion and even simple
models result in very complicated mathematical expressions.~\cite{Tamashiro2010} 
Within the mixed dielectric model of Tamashiro and  Constantino,~\cite{Tamashiro2010} we can make 
an estimate of the electrostatic self-energy of an ion located  at  $z=0$,   
$\beta F_{\rm self}(a)= 5.2 \lambda_B/ a$, for the SPC/E water with $\epsilon_{\rm w}=65$.  We can now interpolate
between the different limits to obtain  $F_{\rm self}(z)$ on the interval $[-a,a]$,
\begin{equation}
\beta F_{\rm self}(z)=\frac{3 q^2 \lambda_B}{4 a} \left( 1+5.93 |1-z/a |^{1.8} \right) \,.
\end{equation}
Finally, we relate the $-0.6\,\rm{V}$ surface potential of SPC/E water to a free energy difference across 
the air-water interface.  If we approximate this free energy as a 
discontinuous drop across a point dipole
layer, we obtain the following model for free energy change due to the surface potential as the 
ion moves across the GDS, 
\begin{equation}
\beta F_{\rm pot}(z)=\chi q \left( 1-z/a \right)  \,.
\label{eq:qchi}
\end{equation}
Here $\chi$ will be determined by the value of the electrochemical surface potential.  
Combining all these contributions, we obtain  the total potential for the ion in the SPC/E water,  $F_{\rm spce}(z)$.

For the Stockmayer liquid there is no free energy due to the electrochemical surface potential 
contribution to the solvation free energy.  The cavitational 
energy is more difficult to define because the air-water interface of the Stockmayer liquid is very diffuse.  
However if we adjust the energy scale so that the ion is fully solvated
at one ionic radius from the GDS, the 
cavitation energy can be written as
\begin{equation}
\beta F_{\rm cav}^{\rm stock}(z)=0.0525 a^3 \left( \frac{z}{a}+1\right)^2 \left( 2-\frac{z}{a}\right)
\end{equation}
The electrostatic self-energy of an ion inside a Stockmayer liquid with $\epsilon_{\rm w}=140$ is
\begin{equation}
\beta F_{\rm self}(z)=\frac{3 q^2 \lambda_B}{4 a} \left( 1+12.81 |1-z/a |^{1.85} \right) \,,
\end{equation}
and the total energy $F_{\rm stock}(z)$ is the sum of these two contributions. 

In order to compare the free energy of ion adsorption from the aforementioned classical MD simulations 
to the DCT outlined above, 
we have used restrained molecular dynamics runs where the histogram of the $z$ coordinate
of the ion is 
reconstructed using the weighted histogram analysis method (WHAM).~\cite{Kumar1992,Roux1995}  Specifically, 32 windows spanning the range from
$[-5, 3]$\,\AA\ and $[-6, 2]$\,\AA\ 
relative to the GDS for the SPC/E and Stockmayer fluid, respectively. 
The ion position was restrained
with a harmonic force constant of 42.82\,kcal/(mol\,\AA) and each window was simulated for at least 2\,ns.

\begin{figure}
\includegraphics[width=17.5cm]{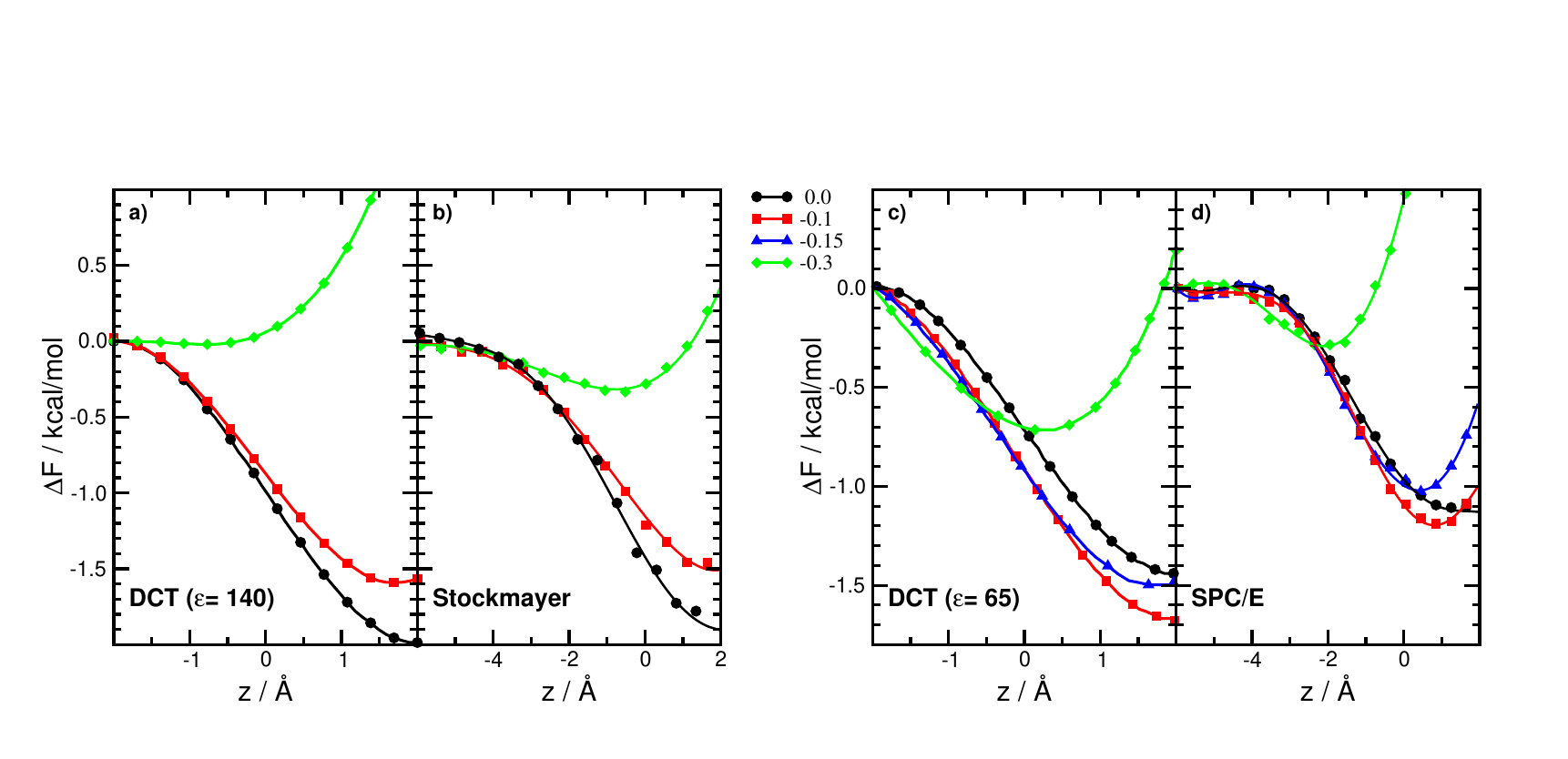}
\caption{%
The potential of mean force of transferring a hard-sphere anion with varying charge across the liquid-air interface 
(negative $z$ bulk, positive air) 
by molecular simulations
in a Stockmayer (b) and SPC/E (d) fluid compared to DCT results for ions with fixed charge 
with $\epsilon=140$ and $\epsilon=65$, respectively.   Each
curve represents the charge state of the hard-sphere. The excellent qualitative
agreement with DCT is apparent in both cases indicating that DCT theory presented here for hard charged sphere adsorption 
does contain the essential physics to model the adsorption seen in our molecular simulations.
The Stockmayer fluid is modeled with an electrochemical surface potential of  $\chi = 0$ and the SPC/E water 
employs the value $\chi=5.82$ in ~\ref{eq:qchi}.  Without the electrochemical surface potential the cavity 
should have the lowest free energy of adsorption.
For the SPC/E, it is clear this is not the case, and the subtle effects obtained from molecular simulation 
are within the error bars of
the WHAM method~\cite{Kumar1992,Roux1995} (see Figure 3 in the SI).   
For completeness, we have presented the PMF for $q=-1e$ (a fully charged 
monovalent ion) for comparison with the DCT
model in the SI where similar agreement is obtained.  Thus, it is clear that we can 
model the statistical mechanics of the full range of monovalent charges for hard-sphere adsorption
with the procedure presented in this letter.
}
\label{fig:pmf-spce}
\end{figure}


The main result of our study is presented in \ref{fig:pmf-spce}.  
Here one can see excellent qualitative agreement with simulation and DCT for the case of a fractionally charged
hard sphere in SPC/E water and  the Stockmayer fluid.  
The  comparison between the simulation and the DCT in \ref{fig:pmf-spce} 
in different panels has been made to establish an important point.  The results in \ref{fig:pmf-spce} are not a result of a fit 
to one another, but are the results of two different modeling approaches.  
Due to its technical nature, a detailed explanation 
of the {\it quantitative}  differences between the simulations and the DCT is given in the SI.  

In the case of  
SPC/E water interacting with a hard sphere in the limit of zero charge, the cavity potential for our simulations
is in excellent agreement with the studies of Rajamani {\it et al.}, which were obtained
using Widom insertion techniques.~\cite{Rajamani2005}  It should also be pointed out that in the DCT study the 
cavitation potential given in Eq.(1) is fit to the simulation 
results of Rajamani {\it et al.},~\cite{Rajamani2005} not
the simulation results presented here (see the SI 
for an explanation of these differences).
Without the 
inclusion of the electrochemical surface potential, the DCT for fixed charge ions dictates that the {\it uncharged}
cavity PMF should produce the 
lowest free energy in the charged cavity series.  In the case 
of the SPC/E simulations, this is clearly not the case.  
Our study shows that indeed small charges adsorb more than the 
uncharged cavity. These, small but real effects can only be accounted for through DCT by
the inclusion of an electrochemical surface potential of $-0.3\,{\rm V}$ corresponding to
$\chi=5.82$ in Eq.(3), or $10k_{\rm B}T$, which is roughly half of 
the computed surface potential for SPC/E water.

It is important to recall that DCT theory assumes, in general,
that the two terms that contribute to the solvation free energy are associated with the
formation of a cavity followed by
the charging process.  The latter is assumed to be completely determined by 
dielectric response.  From the point of view of hard-spheres in a dielectric, the 
asymmetry of cations and anions in the adsorption at the air-water interface is due only 
to the presence of the electrochemical surface potential.  Recent work by  
Arslanargin and Beck has described this term in the context of molecular simulation
as coming from both near and far-field electrostatic contributions.~\cite{Arslanargin2012}  
The near-field is due to the
local arrangements of water molecules around the ion, and the far-field to
the electrostatic potential (surface potential) of the air-water interface.  
Both of these contributions are contained in 
in ~\ref{eq:qchi}, which gives a  free energy gain of $10k_{\rm B}T$, due
to the electrochemical surface potential
of $-0.3\,{\rm V}$, in excellent agreement with the value obtained
by Arslanargin and Beck.~\cite{Arslanargin2012}

When comparing the DCT model to the simulation of a charged hard sphere in a Stockmayer fluid
a very different picture emerges.  Again, we see excellent agreement between the DCT and
our simulations.
However, we find that the {\it uncharged cavity} indeed has
the most negative adsorption free energy, as expected. This is in stark contrast to the PMFs 
obtained for the SPC/E water model, in which the uncharged cavity does not exhibit the lowest free energy.
Moreover, we have verified that cations and 
anions give identical free energy profiles as dictated from the symmetry of the 
Stockmayer fluid.  Thus, we have provided direct evidence through molecular simulation 
that the principles for anion adsorption  are different for the Stockmayer fluid. 
Only fully charged monovalent anions of unrealistic size will be able to adsorb to
the air-water interface.

Horinek and co-workers have fit a point charge, soft-sphere interaction potentials
to yield PMF that reproduce  the experimentally measured surface tensions as a function of 
electrolyte concentration.~\cite{Horinek2009} 
This, however comes with a price since other thermodynamic and structural properties will not
be well described by modifications of the interaction potential.~\cite{Horinek2009} 
On the other hand, it has been shown using {\it ab initio} based interaction potentials that
we recover the experimentally determined local solvation structure around the anion.~\cite{Fulton2010}
Therefore, our present work suggests that the small adsorption 
predicted by Horinek and co-workers is due to the {\it electrochemical surface potential} of SPC/E water.~\cite{Arslanargin2012} 
In other words, the electrochemical surface potential is implicitly contained
in their simulated PMF, while in our treatment it appears explicitly, {\it i.e.,} it is 
separated from the cavity and electrostatic self-energy contributions to the free energy.
Thus, we conclude that  the driving force for adsorption of large halide anions to the interface in 
point charge models of ions and water arises from different principles than in our {\it ab initio} MD simulation 
and PA-DCT theory.
For example, we assert that fully charged,  non-polarizable anions of realistic size will only adsorb if they couple to 
free energy at the air-water interface corresponding to a electrochemical surface potential 
of $-0.3\,\rm{V}$.~\cite{Arslanargin2012} 
In contrast, DFT and PA-DCT suggest it is only cavitation, aided by polarizability, that 
drives large anions to the air-water interface.

It is clear that future research needs to be focused on the role of the 
{$+3\,\rm{V}$ {\it ab initio} derived surface potential and how it 
influences the free energy profile for an ion 
in the vicinity of the vapor-liquid interface. 
There have been other independent
studies that have speculated as to why an ion does not feel the {\rm full} surface potential
given by point charge models~\cite{Harder2008,Kathmann2011,Arslanargin2012}.  
These studies give hints as to why it could be possible for an ion in a quantum
mechanical study to feel a very small and possibly negligible surface potential.
Verification of these ideas with {\it ab initio} calculations
is a subject for future research where, once again, proper simulation protocols must be 
developed in order to isolate the
different contributions to the free energy of adsorption of ions to the air-water interface.

We have studied the driving forces for the adsorption of ions to the air-water interface using both
molecular simulation and DCT for non-polarizable ions.  Within the class of partial 
charge water models ({\it e.g., SPC/E}),
the free energy of adsorption for non-polarizable, hard sphere ions has two contributions:
cavitation, and the negative electrochemical surface potential of water.  The microscopic origin of the cavitation
energy is the perturbation to the hydrogen bond network produced when the ion is inside water. 
When the ion moves across the interface, the perturbation to the hydrogen bond network decreases,
and so does the cavitation energy.
In addition, the classical water models give rise to a negative free energy of adsorption that is associated with
the electrochemical surface potential, which has been shown elsewhere to have both near and far-field contributions.~\cite{Arslanargin2012}
Our calculations suggests that the free energy due to the surface potential
that an ion experiences is approximately half of
the $-0.6\,\rm{V}$ average electrostatic surface potential of SPC/E water, that amounts to a significant
driving force for moving the ion across the interface.
However, moving a hard (non-polarizable) ion completely into the low dielectric of air results in a huge electrostatic self-energy
penalty that overwhelms the favorable contributions from the cavitational energy and the electrostatic surface potential. 
Simulations of the Stockmayer fluid, which by symmetry does
not have a contribution from the electrochemical surface potential, 
have further emphasized the fundamental role played by 
the electrostatic surface potential in determining the
driving force for ion adsorption within classical water models.  

We anticipate that the effect of the electrochemical surface potential of empirical
water models utilizing polarizable force fields 
will yield similar effects on ionic adsorption.  
It has been shown that the computed electrochemical surface potential for water
obtained from simulations using a polarizable force field
is similar to those obtained using non-polarizable point charge models.~\cite{Wick2006} 
Additionally, as argued in many 
recent studies, the currently
available polarizable force fields tend to over-polarize even under {\rm bulk} conditions.~\cite{Wick2009a,Rogers2010,Fulton2010}  
There have been attempts to remedy this problem using different methods for
screening the multipole electrostatics that yield dipole moments in good agreement
with density functional theory studies.~\cite{Wick2006}  However, even these modified models will  
predict too much adsorption ($2 k_{\rm B}T$), leading to an incorrect,
negative surface tension increment for solutions of large halides.  

At this point, it is important to stress that the $-0.6\,\rm{V}$ surface potential of classical water
models is likely due to the nature of a point charge model in the vicinity of a broken symmetry.  
This has been  argued in earlier work~\cite{Wilson1988,Kathmann2008,Kathmann2009,Kathmann2011} where it is shown that, as 
long as the width of the electron
density is larger than that of the positive nuclear charge density, then the mean-inner potential,
or surface potential, will be positive.  
Thus, the real average electrostatic potential across the air-water interface as calculated using the {\it ab initio}
simulations is $+3\rm{V}$ (again, note the difference in sign). In Ref.~\cite{Kathmann2011} it was determined that this 
average potential can be probed by high energy electron holography measurements, but does
not translate directly into the free energy felt by an anion.
Our previous {\it ab initio} study
suggests ions feel a free energy due to the electrochemical surface potential that is small and possibly negligible.\cite{Baer2011}  

Insofar as charged hard-spheres in dielectric media are concerned, it would seem that we have
a way of decomposing the interactions that drive ions to interfaces using molecular
simulation and DCT.  Unfortunately, this simplified
view does not capture the degree of complexity that exists in real systems modeled either by
sophisticated  interaction potentials derived empirically or by quantum mechanics. 
Moreover, in the case of soft-spheres ({\it e.g.} Lennard-Jones interactions) it is much more
difficult to make contact with DCT because one cannot easily separate the cavitation free energy
from the dielectric response.  Despite this complication, similar conclusions concerning the role of the electrochemical
surface potential in driving ions to interfaces were obtained in a recent study of ion adsorption using the soft sphere potentials
of Horinek {\it et al}.~\cite{Arslanargin2012}

Clearly, if the  electrochemical surface potential is directly related to
the surface potential of water (say, $+3\,\rm{V}$, as computed by DFT) it should strongly favor anion {\it repulsion} 
from the surface, and not adsorption as predicted by the classical 
water models. The relationship between the surface potential (which can be probed by high energy electrons)
to the electrochemical surface potential that an ion would feel is still not apparent.  
In the present study, we have shown that we can  isolate the 
free energy contribution due to an electrochemical surface potential in the PMF for the
adsorption of a hard sphere ion to the air-water interface.
The neglect of a free energy due to the electrochemical surface 
potential in the PA-DCT leads to excellent agreement with surface
tension measurements and the PMF for iodide adsorption obtained using DFT.~\cite{Baer2011}
These interesting results suggest that the real electrochemical surface
potential that an ion feels is model dependent and is likely much 
smaller ($\sim\pm10\,{\rm mV}$) than that obtained in simulations
employing classical water models.
Given that we can achieve near quantitative agreement between molecular
simulation and DCT theory for fixed charge ions for the models employed 
in the present investigation, we are confident
that similar connections can be made with {\it ab initio} simulation
and PA-DCT. 

%
\begin{acknowledgement}
It is a pleasure to thank Prof. Tom Beck for sharing preprints 
of his work and for candid discussions on this subject.  
We also acknowledge insightful conversations with Greg
Schenter and Shawn Kathmann.
This work was supported by National Science Foundation grant CHE-0431312
and by the U.S.\ Department of Energy`s (DOE) 
Office of Basic Energy Sciences, Division of Chemical Sciences, 
Geosciences and Biosciences. Pacific Northwest National Laboratory 
(PNNL) is operated for the Department of Energy by Battelle.
The calculations presented herein were performed with PNNLs Institutional Computing.
MDB is grateful for the support of the Linus Pauling Distinguished 
Postdoctoral Fellowship Program at PNNL. Y.L. would like to acknowledge 
partial support by the CNPq, FAPERGS, INCT-FCx, and by the US-AFOSR 
under the grant FA9550-09-1-0283.
\end{acknowledgement}

\begin{suppinfo}
\end{suppinfo}

\newpage

\bibliography{bib-new}
\end{document}